\documentstyle[12pt]{article}
\renewcommand
\baselinestretch{1.4}

\def\beq{\begin{equation}}
\def\brr{\begin{array}}
\def\err{\end{array}}
\def\eeq{\end{equation}}
\def\bea{\begin{eqnarray}}
\def\eea{\end{eqnarray}}
\def\ben{\begin{enumerate}}
\def\een{\end{enumerate}}
\def\Nat{\mbox{\bf N}\, }
\def\Real{\mbox{\bf R}\, }
\def\Comp{\mbox{\bf C}\, }
\def\bs{\bigskip}
\def\tr{\mbox{Tr}\, }
\def\ni{\noindent}

\def\wh{\widehat}

\def\nn{\nonumber}
\def\ms{\medskip}

\def\Re{\mbox{Re}\, }

\def\dsp{\displaystyle}

\newcommand{\ii}{\infty}                         
\newcommand{\fr}[2]{\mbox{$\frac{#1}{#2}$}}      
\newcommand{\Tr}{\,\mbox{Tr}\,}                  
\renewcommand{\Re}{\,\mbox{Re}\,}                
\newcommand{\al}{\alpha}
\newcommand{\be}{\beta}

\newcommand{\ze}{\zeta}

\newcommand{\Ga}{\Gamma}

\begin{document}

\hfill UB-ECM-PF 93/7


\hfill March 8, 1993

\vspace*{3mm}

\begin{center}

{\LARGE \bf
Mellin transform techniques for zeta-function resummations}

\vspace{4mm}

\renewcommand
\baselinestretch{0.8}

{\sc E. Elizalde}\footnote{E-mail address: eli @
ebubecm1.bitnet}, {\sc S. Leseduarte}, \\
{\it Department E.C.M., Faculty of Physics, University of
Barcelona, \\
Diagonal 647, 08028 Barcelona, Spain}, \\
 and  {\sc S. Zerbini} \\
{\it Department of Physics, University of Trento, 38050 Povo,
Italy} \\
{\it  and I.N.F.N., Gruppo Collegato di Trento}
\ms

\renewcommand
\baselinestretch{1.4}

\vspace{5mm}

{\bf Abstract}

\end{center}

Making use of inverse Mellin transform techniques for analytical
continuation, an elegant proof and an extension of the
zeta function regularization theorem
 is obtained. No series
commutations are involved in the procedure; nevertheless
the result is naturally split into the same three
contributions of very
different nature, i.e. the series of Riemann zeta functions and the
power and
negative exponentially behaved functions, respectively, well known
from the
original proof.
The new theorem deals equally well with elliptic differential
operators
whose spectrum is not explicitly known. Rigorous results
on the asymptoticity of the outcoming series are given, together
with
some specific examples. Exact analytical
formulas, simplifying approximations and numerical estimates  for
the last of the three contributions (the most difficult to handle)
are
obtained. As an application of the method, the summation of the
series which appear in
the analytic computation (for different ranges of temperature) of
the
partition function of the string ---basic in order to ascertain if
QCD is some limit of a string theory--- is performed.



\newpage

\section{Introduction}

Zeta-function resummation formulas are essential in the
zeta-function regularization procedure \cite{1}. In particular, they
constitute
the key ingredient for the proof (done in several steps by Weldon,
Actor, Elizalde and
Romeo \cite{2,3}) of the zeta function regularization theorem.
The importance of the final outcome of the theorem ---and also, of
its extension to multiple, generalized Epstein-Hurwitz series
with arbitrary exponents \cite{3}--- has been stressed by several
authors
\cite{4}. Specifically, it is necessary for actual, practical
applications to estimate or put a bound to the error one introduces
by
neglecting the non-polynomical terms which arise in the series
commutation process. These terms have proven to be very difficult
to
handle and, due to the fact that they are (very) small in some
cases,  usually they have just been dropped off from the final
formulas.
To put an end to such an unpleasant situation is one of the
purposes
of this paper.

Another motivation is to present a completely different method for
the derivation of these additional contributions, which comes about
from
a very elegant approach that has its roots in an elaborate
admixture of the Mellin transform technique and the heat kernel
method. Our new procedure yields a very convenient, closed
expression for
the zeta function corresponding to very general elliptic operators
in
terms
of complex integrals over movable vertical lines in the complex
plane. The final result is the same as the one painstakingly
obtained {\it via} the original method
of the series commutation ---with the great advantage that the
proof of
asymptoticity of the resulting series turns to be now immediate and
also, that the calculation of bounds to the additional term is
considerably easied. Moreover, cast in this form the theorem admits
a quite
natural generalization to the case where the spectrum of the
operator is not known.

In the next section of the paper, we describe the Mellin transform
method as applied to generalized heat kernel operators. The method
is extended in section 3 to the case of
 trace formulas of generalized zeta-function type.
 In section
4 some mathematical theorems on asymptoticity of series, basic for
practical applications of the procedure in rather general
situations
are presented.
 In section 5 we obtain some initial (but
already non-trivial) results on the additional term by making
convenient use of the Poisson summation formula.
Specific numerical evaluations and an analytical
bound  on  the elusive additional term which appears
as a byproduct of, e.g., the series commutation procedure,
are obtained in section 6, together with some convenient integral
expressions for this term.
 As a particular
example of the usefulness of  the whole
procedure, in section 7 we obtain exact expressions for the
analytical
continuation of highly non-trivial, zeta-type series which appear
in the calculation of
the
partition functions of strings and membranes. This issue happens to
be
essential in deciding if QCD can (or cannot) in fact be described
as a
(super)string (or membrane) theory. Finally, section 8 is devoted
to conclusions. \bs

\section{Inverse Mellin transform method for
generalized heat kernel operators}

We shall here rederive the (by now well known) zeta-function
resummation formulas ---using the Mellin transform
techniques that have been employed for obtaining
high-temperature expansions,
 both in field theory as well as in
string theory \cite{5}---
in a very straightforward way,

Let us consider a non-negative elliptic operator $A$
of order $d$, defined on a compact
$n$-dimensional manifold.
Let  $\{ \lambda_i,\phi_i \}$ be the
spectral resolution of $A$. (The explicit knowledge of the spectral
resolution of the operator will
not be basic to our method, and we shall make only a very limited,
circumstancial use of it.) \,
In what follows we shall be interested in the quantity
\begin{equation}
F_t(s)=A^{-s} e^{-tA},
\label{1}
\end{equation}
which can be interpreted as a generalized heat kernel operator. It
satifies the functional differential equation
\begin{equation}
\partial_s F_t(s)+F_t(s-1)=0
\label{2}
\end{equation}
and, from the operator identity
\begin{equation}
e^{-tA}=\frac{1}{2\pi i}\int_{\Re \dsp z=c} dz \, \Gamma(z) \,
(tA)^{-z},
\label{3}
\end{equation}
one obtains the complex integral representation
\begin{equation}
F_t(s)=\frac{1}{2\pi i}\int_{\Re \dsp z=c}dz \, \Gamma(z) \,
t^{-z}A^{-s-z}.
\label{4}
\end{equation}

Since $A$  is assumed to be elliptic, the diagonal part of its heat
kernel
as well as the kernel of the related zeta function exist. A theorem
of Seeley yields
\begin{equation}
\zeta_A(s+z)(x)\simeq \frac{1}{\Gamma(s+z)}\left[\sum_r
\frac{K_r(x)}{s+z+\frac{r-n}{d}}+
\hat{J}(s+z,x)  \right]
\label{5}
\end{equation}
(where $\simeq$ is the symbol for asymptotical equivalence), being
$K_r(x)$ the heat kernel expansion coefficients and
$\hat{J}$
the analytical part of $\zeta_A$, which is not known
explicitly, in general. As a result of all the preceding
expressions,
one arrives at
\begin{equation}
F_t(s,x)\simeq \frac{1}{2\pi i}\int_{\Re \dsp z=c}dz \,
t^{-z}\frac{\Gamma(z)}
{\Gamma(s+z)} \left[\sum_r \frac{K_r(x)}{s+z+\frac{r-n}{d}}+
\hat{J}(s+z,x) \right].
\label{6}
\end{equation}
The corresponding trace formula can be obtained by integrating
over $x$, with
the result
\begin{equation}
\tr F_t(s) \simeq \sum_i \lambda_i^{-s} e^{-t\lambda_i}=
\frac{1}{2\pi i}\int_{\Re \dsp z=c}dz \, t^{-z}\frac{\Gamma(z)}
{\Gamma(s+z)} \left[\sum_r \frac{K_r}{s+z+\frac{r-n}{d}}+
\hat{J}(s+z) \right].
\label{7}
\end{equation}
In the rest of the section  we will employ this form of
complex integral representation,
with the understanding that all the conclusion we shall draw using
this method can be also validated for
the corresponding non-integrated quantities.
In order to obtain the so-called resummation formulas (see Actor,
Weldon, Elizalde and Romeo \cite{2,3}) one needs to perfom
a
very accurate application of the residues theorem. For a generic
$s$, we encounter simple poles at $ z=-k$, $k=0,1,2, \ldots$, and
at $z=\frac{n-r}{d}-s$ in the expression above.
Furthermore, in general there will be a non-zero contribution
coming from the integral evaluated
along the semicircumference at infinity (see Elizalde and Romeo,
refs. \cite{2,3}).
As a result, we can decompose $F_t (s)$ into three parts, very
different in nature,
\begin{equation}
F_t(s)\simeq F_t^{(1)}(s)+ F_t^{(2)}(s)+ F_t^{(3)}(s),
\label{8}
\end{equation}
which turn out to be, in our language,
\begin{equation}
F_t^{(1)}(s) = \sum_k \frac{(-t)^k}{k!} \, \zeta_A(s-k),
\label{9}
\end{equation}
\begin{equation}
F_t^{(2)}(s) = \sum_r
\frac{\Gamma(\frac{n-r}{d}-s)}{\Gamma(\frac{n-r}{d})} \,
t^{s+\frac{r-n}{d}}K_r
\label{10}
\end{equation}
and
\begin{equation}
F_t^{(3)}(s) = \frac{1}{2\pi i}\int_C  dz \, \Gamma(z) \,
t^{-z}\zeta_A(s+z),
\label{11}
\end{equation}
 $C$ being a convenient contour on the complex plane.
Some remarks are in order. First, these results are in complete
agreement with the ones originally obtained by one of the authors
(see also the recent paper by Actor \cite{6}) for specific
situations. However, in our new
approach, we explicitly see how the heat kernel coefficients $K_r$
enter the game, in the general case considered from the
begining (eq. (\ref{1})).
These are, in principle, computable quantities (for a recent
reference see Branson and Gilkey \cite{7}), in terms of which the
characteristics
of an elliptic operator are always given. The analytic part of the
final result
---which is generally unknown--- enters into the first and
third contributions.
In some cases (see, for instance,
Elizalde and Romeo \cite{2}), it can be explicitly evaluated. In
particular, it is interesting to consider also the case of
hyperbolic compact manifolds, where the analytic part can be
related to a
 zeta function of Selberg type (Bytsenko et al. \cite{5}), even though
yet in this case the spectrum is usually not known. Also, we
 should notice that a similar result has been obtained by
Avramidi in a particular case \cite{8}.

A very important  remark concerns the {\it range}
 of the sums in eqs.
(9)-(11). In previous works \cite{2,3,4} this range was formally
taken to be infinite:
the sums were series and the integration contour $C$ in $F^{(3)}$
was at infinite distance from the origin. Then these expressions
can only have a formal character: on the one hand
 the series converge only in
{\it very specific cases} and, on the other,
 the contribution of the contour is usually
 infinite. This will {\it not} be any more our philosophy in the
present paper. In the following sections we shall prove that (under
very general conditions) the series in (9)-(11) are asymptotic;
thus, we will always consider them as cut after the most favourable
term (yielding always {\it finite} sums), and the contour $C$ will
be a vertical line at a constant finite abscissa $c$ (i.e., Re
$z=c$), together with
 the two corresponding horizontal segments at infinity.
This is the central part of the full method in this paper, and
 must be kept always in mind, from now on, when considering
 expressions like
 (9)-(11). Rigorous specifications
 on this point will be given in the
next sections.
\bs

\section{Extension
 to trace formulas of generalized zeta-function type}

The above technique can be extended to trace formulas of the kind
\begin{equation}
F(s,t)=\sum_i \lambda_i^{-s} f(t\lambda_i),
\label{12}
\end{equation}
provided $f(y)$ admits an invertible Mellin transform, namely
\begin{equation}
f(y)=\frac{1}{2\pi i}\int_{\Re \dsp z=c}dz \, y^{-z}M[f](z),
\label{13}
\end{equation}
with
\begin{equation}
M[f](z)=\int_0^\infty dy \, y^{z-1}f(y).
\label{14}
\end{equation}

Being more precise, it is assumed that the function $f$ is
integrable
over the open interval $(t_1,t_2)$ provided that $0<t_1<t_2$.
Defining
\beq
\alpha \equiv \inf \left\{ \alpha^*\in \Real | f(0^+)={\cal O}
(t^{-\alpha^*}) \right\}, \ \ \ \ \beta \equiv \sup \left\{
\beta^*\in \Real | f(+\infty ) {\cal O} (t^{-\beta^*}) \right\},
\eeq
we shall also consider that $\alpha <\beta$. This is no
restriction
since if, on the contrary, $-\infty < \beta \leq \alpha < \infty$,
one just needs to define (for $R>0$) $f_1(t)\equiv \theta (R-t)
f(t)$, $f_2(t)\equiv \theta (t-R) f(t)$ and $F_j(s,t)=\sum_i
\lambda_i^{-s} f_j(t\lambda_i)$, $j=1,2$, and then consider each
$F_j$ separately (with $\alpha_1 = \alpha$, $\beta_1=+\infty$, and
$\alpha_2 =-\infty$, $\beta_2=+\beta$).

In general, there will exist a value $s_0\in \Real$ such that for
Re
$s>s_0$ the series $\sum_i \lambda_i^{-s}$ will be absolutely
convergent and for Re $s<s_0$ it will be not. Then, a sufficient
condition in order that (\ref{12}) makes sense is  Re
$s+\beta>s_0$. In this case,
\beq
F(s,t)=\sum_i \lambda_i^{-s} \frac{1}{2\pi i}\int_{\Re \dsp z=c}dz
\, (\lambda_i t)^{-z}M[f](z),
\eeq
with $c\in (\alpha, \beta)$ and Re $(s) +c >s_0$. We take
 $M[f] (c+iy)\in L^1 (-\infty <y <\infty)$, so that,
finally, making use of all these equations, we can write
\begin{equation}
F(s,t)=\frac{1}{2\pi i}\int_{\Re \dsp z=c}dz \, t^{-z}M[f](z) \,
\zeta_A(s+z).
\label{15}
\end{equation}

Now, we must suppose that one can solve the problem of the
analytic
continuation of the functions $\zeta_A (z)$ and $M[f](z)$ to the
left
in Re $z$. Under very general conditions, the function $M[f]$ can
be
extended as a meromorphic function, but for a general $\zeta_A$
this is
not so easy to do in practice. In general one must calculate all
the
Seeley-De
Witt coefficients of $A$ and use them to infer the meromorphic
behavior
of $\zeta_A$ (in the second example below we show an
alternative way to extend $\zeta_A$ using our knowledge of the
$\lambda_i$ only).  We also need to know the behavior of  $\zeta_A
(z)$ and $M[f](z)$ for $|\mbox{Im} \ z| \rightarrow \infty$.

As before, for a generic $s$, we can split
\begin{equation}
F(s,t)=F^{(1)}(s,t)+ F^{(2)}(s,t)+ F^{(3)}(s,t),
\label{16}
\end{equation}
where now
\begin{equation}
F^{(1)}(s,t)= \sum_k  \mbox{Res} \, M[f](z_k) \,
t^{-z_k}\zeta_A(s+z_k),
\label{17}
\end{equation}
\begin{equation}
F^{(2)}(s,t)=\sum_{r}
\frac{M[f](\frac{n-r}{d}-s)}{\Gamma(\frac{n-r}{d})}
\, t^{s+\frac{r-n}{d}}K_r
\label{18}
\end{equation}
and
\begin{equation}
F^{(3)}(s,t)=\frac{1}{2\pi i}\int_C  dz \,
M[f](z) \, t^{-z}\zeta_A(s+z),
\label{19}
\end{equation}
being $z_k$ all the (simple) poles of $M[f](z)$.
If double poles occur, logarithmic terms in $t$ show up. This may
happen for particular values of $s$, once the other free parameters
are held fixed.

Let us now consider some illustrative examples.
\ms

\ni{\bf Example 1}. For the Riemann zeta function itself, we have
\beq
\zeta (s)= {\cal O} \left( |t|^{p (r)} \ln |t| \right), \ \ \ \
s=r+it,
\label{ze1}
\eeq
with
\beq
p(r) = \left\{ \brr{ll} \frac{1}{2}- r, & -K \leq r \leq -\delta,
\\
\frac{1}{2}- r, & -\delta \leq r \leq 0, \\
\frac{1}{2}- r, & 0 \leq r \leq \delta, \\
1- r, & \delta \leq r \leq 1-\delta, \\
1, & 1-\delta \leq r \leq 1, \\
0, & 1 \leq r \leq 1+ \delta, \\
0, &  r \geq 1+\delta, \err \right.
\label{ze2}
\eeq
where $K>0$ and $\delta >0$ (a standard choice is
$\delta <1$ and $K$ arbitrarily big). Actually, $\ln |t|$ can be
supressed in the first, fourth and seventh intervals. It is
remarkable
that eq. (\ref{ze1}) is valid uniformily in each of the intervals
of
(\ref{ze2}).
\ms

\ni{\bf Example 2}. Take $\lambda_n=n$ and add a degeneration
$g(n)$ (being $g$ an analytical function with good behavior in
order to fulfill what follows), i.e.
\beq
\zeta_g (z)\equiv \sum_{n=1}^{\infty} \frac{g(n)}{n^z} = -
\frac{i}{2\pi} \int_{\cal C} \ln [\sin (\pi t)] \, \frac{d}{dt}
(t^{-
z}g) \, dt,
\eeq
being $\cal C$ the contour of the sector of the complex plane
defined by $|\arg z| \leq \theta_0$ and $|z|\geq \epsilon$,
$0<\epsilon <1$. Decomposing $\cal C$ into its upper and lower
parts,
$\cal C^{\pm}$, and assuming $g$ to be analytical on the sector, we
obtain
\beq
\zeta_g (z)= -\int_{\epsilon}^{\infty} d\rho \, \rho
\frac{d}{d\rho} (\rho^{-z}g) - \frac{1}{2} \epsilon^{-z}
g(\epsilon) -\frac{i}{2\pi} \sum_{\pm} \int_{\cal C^{\pm}} dt \,
\ln \left( 1-e^{\pm 2\pi i t} \right) \frac{d}{dt}  (t^{-z}g).
\eeq
In general, the second and third terms on  the r.h.s. are integer
functions of $z$, and only the first one needs to be continued to
the
left in Re $z$. This can be done by applying the
Mellin transform techniques.
\bs

\section{A more explicit study of the additional term}

Of the three terms which appear in each of the splittings (\ref{8})
and
(\ref{16}), the third one is the most difficult to tackle. In fact,
it was completely overlooked in the first formulations of the zeta
function regularization theorem, what led to several erroneous
results.  In the  procedure above it has shown up in a nice  and
elegant
way. However, its apparently very simple form is somewhat
deceiving.
In fact, in its primitive form \cite{2,3}, it required an evaluation of
the integrand
---which involves the zeta function of the elliptic operator--- all
the
way along a contour of {\it infinite} radius. We shall here show
explicitly how to
obtain an alternative asymptotic expansion of expressions
(\ref{8}) and
(\ref{16}) in which the residual term consists of a complex
integration on a vertical line of constant, finite abscissa. We
start by proving the following theorem.
\ms

\ni{\bf Theorem 1}. Let $\theta_0 >0$ and $h$ be an analytic
function
in the domain defined by $z\neq 0$ and $|\arg z | \leq \theta_0$,
and such that
\beq
h(z) \sim \sum_{n=0}^{\infty} a_n z^{\beta_n}, \ \ \ \ |z|
\rightarrow 0,
\label{t11}
\eeq
and (only to simplify the discussion, by putting $\beta = \infty$)
\beq
h(z) <{\cal O} \left( z^{-R}\right), \ \ \ \ \forall R \in \Real,
\ \ \ \ |z| \rightarrow \infty,
\label{t12}
\eeq
where we understand that Re $\beta_n\leq $ Re $\beta_{n+1}$ and Re
$\beta_n \rightarrow \infty$. Define
\beq
Y_r (s, \tau) \equiv \sum_{n=1}^{\infty} \frac{h (n^r \tau)}{n^s},
\label{y1}
\eeq
with $r,\tau >0$ and $s \in \Comp$. Then, $Y_r (s, \tau)$ is an
integer function with respect to the argument $s$ and, denoting as
before the Mellin transform of a function $f$ by $M[f]$, one has
\bea
Y_r (s, \tau) & \sim& \tau^{-(1-s)/r} \left[ \frac{s}{r} M[h]
\left(
\frac{1-s}{r} \right) - M[h]' \left( \frac{1-s}{r}+1 \right)
\right]-
\frac{1}{2} \tau^{s/r} \left[ \frac{s}{r} M[h] \left(- \frac{s}{r}
\right) \right. \nn \\ &-& \left. M[h]' \left(1- \frac{s}{r}
\right)
\right]
+ \sum_{n=0}^{\infty} a_n \zeta (s-\beta_n r) \tau^{\beta_n}.
\label{s1}
\eea

\ni{\bf Proof}. Write
\beq
Y_r (s, \tau) =- \frac{1}{2\pi i} \int_C dz \, \pi \cot (\pi z) \,
z^{-s} h(z^r\tau )= \frac{1}{2\pi i} \int_C dz \, \ln \sin (\pi z)
\, \frac{d}{dz} \left[ z^{-s} h(z^r\tau )\right],
\label{24}
\eeq
$C$ being the circuit made up of the two straight lines at angles
$\pm \theta_0$. From the first hypothesis it turns out that the two
integrals in which (\ref{24}) splits exist for any $z\neq 0$ with
$|\arg
z| \leq \theta_0 -\delta$, $\forall \delta >0$. Thus we deform the
circuit $C$ to $C_1$ corresponding to the lines at angles $\pm
(\theta_0
- \delta)$ and with a small circular arc of radius $\epsilon$ at
the
origin.

Notice now the asymptotic behavior of a function of the following
type
\beq
\int_{\epsilon}^{\infty} d\rho \, \rho^{-s} h(\rho^r\tau) \sim
\frac{\tau^{-(1-s)/r}}{r} \, M[h]  \left( \frac{1-s}{r} \right) -
\sum_{n=0}^{\infty} a_n \, \frac{\tau^{\beta_n}}{r} \,
\frac{\epsilon^{r\beta_n -s+1}}{\beta_n +\frac{1-s}{r}},
\eeq
provided there are no pole coincidences.
Then, defining
\beq
Y_1 \equiv \frac{is}{2\pi} \int_C dz \, \ln \sin (\pi z) \, z^{-s-
1} h(z^r\tau )\equiv Y_1^+ + Y_1^-,
\eeq
corresponding naturally to the two branches of the circuit,
$C_{\pm}$,
the analysis of each of these integrals is quite simple. The second
part
 can in fact be treated as the first, by noticing that
\beq
Y_2 \equiv \frac{r\, \tau}{2\pi i} \int_C dz \, \ln \sin (\pi z) \,
z^{r-s-1} h'(z^r\tau ) =\frac{-r\, \tau}{s-r} Y_{1,r}^{[h
\rightarrow
h']} (s-r, \tau).
\eeq
Collecting everything together, one arrives at the desired
expression (\ref{s1}), valid in principle for any positive integer
$n$ such that $s\neq 1+\beta_n r$ and  $s\neq \beta_n r$.
These cases must be considered specially. Either one starts over
again or, what is better, one proceeds according to the following
hint:
to perform a shift $s=1+\beta_n r+\delta$ or, correspondingly,
$s=\beta_n r+\delta$, and then make use of the well known lemma
\ms

\ni{\bf Lemma 1}. Let $D$ an open domain of {\bf C}, $s_0\in D$,
$D^* =D-\{ s_0\}$, and $\beta_n$ a sequence in {\bf C} such that
 Re $\beta_n\leq $ Re $\beta_{n+1}$ and Re
$\beta_n \rightarrow \infty$. Let $Y(s,\tau)$  analytic as a
function of the first argument for $s\in D$ and $\tau \in (0,
\alpha)$, $\alpha >0$. Suppose that for $\forall s\in D^*$ one has
 $Y(s,\tau)\sim \sum_{n=0}^{\infty} a_n (s) \, \tau^{\beta_n}$,
where the $a_n(s)$ are analytic on $D$, and that this expansion
is valid uniformily for $s\in C$, $C$ being a path contained in $D$
and enclosing $s_0$.

Then it turns out that also  $Y(s_0,\tau)\sim \sum_{n=0}^{\infty}
a_n (s_0) \, \tau^{\beta_n}$. The proof is very simple.
\ms

In order to determine the behavior of $M[f] (x+iy)$ for
$|y|\rightarrow
\infty$ the following theorems are useful
\ms

\ni{\bf Theorem 2}. Let $f\in C^n(0,\infty)$ and supose there
exists $x_0\in \Real$ such that, for $x>x_0$, $\lim_{t\rightarrow
\infty} (td/dt)^p (t^xf)=0$, for $p=0,1,2, \ldots, n$, and that
$t^{-1} f_p$ is absolutely integrable, where $f_p(t,x)\equiv
(td/dt)^p (t^xf)$. Then $M[f](z) ={\cal O} (|y|^{-n})$ for
$y\rightarrow \pm \infty$.

\ni{\bf Proof}. Follows immediately by wishful application of
  Riemann-Lebesgue's lemma.
\ms

\ni{\bf Theorem 3}. Let $f\in C^n(0,\infty)$ with the asymptotic
behavior for $t\rightarrow 0^+$: $f(t) \sim  \sum_{m=0}^{\infty}
p_m t^{a_m}$, with  Re $a_m$ monotonically increasing towards
infinity. Suppose also that asymptotic expansions for
$t\rightarrow
0^+$ for the succesive derivatives of $f$ are obtained by taking
the derivative, term by term, of the above expansion for $f$.
Suppose also that  $\lim_{t\rightarrow \infty} f_p(t,x)=0$,  for
$p=0,1,2, \ldots, n$, and that $t^{-1}f_p$ is absolutely integrable
for $x>-$ Re $a_0$. Then it follows that $M[f] (z)= {\cal O}
(|y|^{-
n})$ for $y\rightarrow \pm \infty$ and for any $x$, that is, in the
whole domain of the function, after its continuation.

\ni{\bf Proof}. Let $\rho \in \Real$ be large enough and $\mu$ such
that
Re
$a_{\mu-1}
< \rho \leq $ Re $a_{\mu}$, where for a positive integer $\delta$
satisfying Re $a_0 +\delta >$ Re $a_{\mu}$, we define
$\sigma_{\rho}
(t) \equiv \exp (-t^{\delta}) \, \sum_{m=0}^{\mu -1} p_m t^{a_m}$,
and $\wh{f} \equiv f- \sigma_{\rho}$. Applying the Theorem 2 to
$\wh{f}$ and noticing that $M[\sigma ]_{\rho} (z) = \sum_{m=0}^{\mu
-
1} (p_m/\delta) \Gamma ((z+a_m)/\delta)$ has exponential behavior
for  $|y|\rightarrow \infty$, from $M[f]= M\wh{f} + M[\sigma
]_{\rho}$
(which gives the continuation to -Re $a_{\mu} <$ Re $z<\beta$) and
the
fact that $\rho$ is arbitrarily big we prove the theorem.
\ms

For $x_0,\theta_0\in \Real$, let us define that a function $h$
belongs to $K$, $h\in K(x_0,\theta_0)$ iff for any $x>x_0$ and
$\epsilon >0$ we have $M[h] (x+iy) = {\cal O} [\exp (-(\theta_0-
\epsilon) |y| )]$, for $|y|\rightarrow \infty$. Also, denote the
sector $S(\theta_0) = \{ t\neq 0 | \, |\arg t| <\theta_0\}$.
\ms

\ni{\bf Theorem 4}. Supose that in the sector  $S(\theta_0)$: (i)
$h$ is analytic; (ii) $h= {\cal O} (t^{\alpha})$, $t\rightarrow
0^+$; and (iii) $h=\exp (-at^{\nu})  \sum_{m=0}^{\infty}
\sum_{n=0}^{N(m)} c_{mn} (\ln t)^n t^{-r_m}$, $t\rightarrow
\infty$, with Re $a\geq 0$, $\nu >0$ and Re $r_m$ monotonically
increasing towards infinity. Then: (i) if $a=0$, $h\in K(-
\mbox{Re} \, \alpha, \theta_0)$; (ii)  if Re $a>0$, $h\in K(-
\mbox{Re} \, \alpha, \theta)$, where $\theta = \min (\theta_0, (\pi
-2|\arg a|)/(2\nu))$.

\ni{\bf Proof}. We only do the first case (the second is
analogous).
Observe that $M[h] (z)$ is analytical for Re $z>-$ Re $\alpha$.
Define
$\theta' =\theta- \epsilon$, $\theta$ as given in the theorem
and $\epsilon >0$; deforming the circuit by an angle $\pm \theta'$
and
making the change of variable $t=r
e^{\pm
i \theta'}$ ($r$, real, will be the new variable of integration)
in the
Mellin transform integral, one easily shows
that $M[h] (z) =  {\cal O} [\exp (-\theta' |y| )]$, for
$|y|\rightarrow
\infty$, $x>$ Re $-\alpha$.
\ms

With a similar strategy as that used in Theorem 3, with  convenient
choices of the function $\sigma_{\rho}$, one can prove results
analogous to Theorem 4 (so called {\it theorems of exponential
decrease})
valid for a range of $x$ which moves towards the left. The only
proviso
is that $h$ have a convenient asymptotic expansion around 0.

Also to be noticed is the fact that in order to apply these
theorems, with the aim of displacing the contour of integration in
(\ref{15}),
one must make sure that the conclusions of Theorems 2-4 are valid
{\it uniformily} on any segment of the $x$ variable along which we
want to displace the integration contour. If we can perform the
translation from $x=c$ to $x=c'$, $c'<c$, and we check that
$\zeta_A (s+c'+iy) M[f] (c'+iy)$ is absolutely integrable, then
\beq
F(s,t)= \sum_{c'< \mbox{Re} \, z<c} \mbox{Res} \ \left[ \zeta_A
(s+z) t^{-z} M[f](z) \right] + \frac{t^{-c'}}{2\pi i} \int_{\Re
\dsp z=c'}dz \, \zeta_A(s+z) t^{-iy}M[f](z).
\label{fstf}
\eeq

To finish this section, let us consider a different example which
cannot be resolved by direct application of Theorem 4 above (since
it
corresponds to $a\neq 0$ and Re $a=0$).
\ms

\ni{\bf Example 3}. Let
\beq
F(s,t)=\sum_{n=1}^{\infty} \frac{J_{\mu} (nt)}{n^s},
\eeq
that is, $\lambda_n=n$, $f=J_{\mu}$, $\alpha=-\mu$, $\beta =1/2$.
We have
\beq
M[J]_{\mu} (z)= \frac{2^{z-1}\Gamma \left( \frac{z+\mu}{2}
\right)}{\Gamma \left( \frac{\mu-z+2}{2} \right)},
\label{mga}
\eeq
with $s_0=1$, Re $s>1/2$ (in this case the domain of $s$ can be
enlarged), and $c> 1-$ Re $s <1/2$. $M[J]_{\mu}$ has poles for $z=-
\mu+2n$, $n\in \Nat$, with residues $(-1)^n 2^{-\mu-2n}/[n!\, (\mu
+n)!]$, while $\zeta (s+z)$ has a pole at $z=-s+1$. From
(\ref{mga}) and from the behavior of the gamma function we know
that in any closed interval of $x=$ Re $z$ we get the behavior
\beq
M[J]_{\mu} (z)= {\cal O} (|y|^{x-1/2}),
\ \ \ \  |y|\rightarrow \infty,
\eeq
uniformily in $x$. We have Re $s =1/2+\epsilon_1$, $\epsilon_1>0$,
and choose $c=1/2-\epsilon_2$, with $0<\epsilon_2<\epsilon_1$, and
any $\delta$, $0<\delta< \min \{ 1/2, \epsilon_1-\epsilon_2\}$.

Consider now the following, segmentwise, displacement
\beq
G(z)=M[J]_{\mu} (z) \, \zeta (s+z).
\eeq
Then it turns out that
\ben
\item for $1/2+\delta-\epsilon_1 \leq x \leq 1/2-\epsilon_2$,
$|M[J]_{\mu} (z)| = {\cal O} (|y|^{-1/2-\epsilon_2})$ and
 $|\zeta (s+z)| = {\cal O} (1)$, therefore
\beq
|G (z)| = {\cal O} (|y|^{-1/2-\epsilon_2};
\eeq
\item for $1/2-\epsilon_1 \leq x \leq 1/2+\delta-\epsilon_1$,
$|M[J]_{\mu} (z)| = {\cal O} (|y|^{-1/2-\epsilon_1})$ and
 $|\zeta (s+z)| = {\cal O} (\ln |y|)$, therefore
\beq
|G (z)| = {\cal O} (|y|^{-1/2-\epsilon_1}\ln |y|);
\eeq
\item for $-1/2+\delta-\epsilon_1 \leq x \leq 1/2-\epsilon_1$,
$|M[J]_{\mu} (z)| = {\cal O} (|y|^{-3/2+\delta-\epsilon_1})$ and
 $|\zeta (s+z)| = {\cal O} (|y|^{1-\delta}\ln |y|)$, therefore
\beq
|G (z)| = {\cal O} (|y|^{-1/2-\epsilon_1}\ln |y|);
\eeq
\item for $-1/2-\epsilon_1 \leq x \leq -1/2+\delta-\epsilon_1$,
$|M[J]_{\mu} (z)| = {\cal O} (|y|^{-3/2+\delta-\epsilon_1})$ and
 $|\zeta (s+z)| = {\cal O} (|y|^{1/2})$, therefore
\beq
|G (z)| = {\cal O} (|y|^{-1+\delta-\epsilon_1});
\eeq
\item finally, for $-K-1/2-\epsilon_1 \leq x \leq -1/2-\epsilon_1$,
$|M[J]_{\mu} (z)| = {\cal O} (|y|^{-x-1})$ and
 $|\zeta (s+z)| = {\cal O} (|y|^{-x-\epsilon_1})$, therefore
\beq
|G (z)| = {\cal O} (|y|^{-1-\epsilon_1}),
\eeq
where $K$ is arbitrarily large.
\een
\bs

\section{The additional term and the Poisson resummation formula }

We  would like first to examine a particular but important case, as
an illustration of the difficulties associated with the
determination
 of the additional term. Such case will be considered also  in the
next
section but under a different point of view.

Let us consider the selfadjoint operator $A=H^{\beta}$,
where $H$ is the Dirichlet Laplacian on $[0,1]$ with eigenvalues
$\lambda_n=n^2$. Putting $ 2\beta=\alpha$, we can write
\beq
\zeta_A(z)=\sum_{n=1}^\infty n^{-\alpha z}=\zeta (\alpha z).
\eeq
The related heat kernel trace reads  ($t>0$)
\beq
\Tr e^{-tA}=\sum_n e^{-n^\al t}\equiv S_\al (t).
\eeq
Making use of the Mellin representation discussed in Sec. 2, we get
\bea
S_\al(t)&=&\frac{1}{2\pi i}\int_C dz t^{-z} \ze( \al z)\Ga(z)=
\frac{\Ga(\fr{1}{\al})}{\al t^{\fr{1}{\al}}}+\sum_{k=0}^\ii
\frac{(-t)^k}{k!}\ze(-\al k)+
\Delta_{\al}(t)\nn\\
&=& \frac{\Ga(\fr{1}{\al})}{\al t^{\fr{1}{\al}}}
-\frac{1}{2}+\sum_{k=1}^\ii \frac{(-t)^k}{k!}\ze(-\al k)+
\Delta_{\al}(t),
\label{46}
\eea
where the power series in $t$ will converge for $|t|<b$ (for some
$b$, abscissa of convergence). Then we
can analytically continue  to  the remaining values of $t$.
The quantity $S_\al(t)$ may be evaluated by making use of the
Poisson formula, which states that for
$f(x)=f(-x)$ and $f(x) \in L_1$, the following equation holds:
\beq
\sum_{n=1}^\ii f(n)=-\frac{1}{2}f(0)+\int_0^\ii dxf(x)+
2 \sum_{n=1}^\ii \int_{0}^\ii dx f(x) \cos 2\pi n x.
\eeq
Let us consider the function
\beq
f(x)=e^{-t|x|^\al}.
\eeq
An elementary computation permits to write
\beq
S_{\al}(t)=-\frac{1}{2} + \frac{\Ga(\fr{1}{\al})}{\al
t^{\fr{1}{\al}}}
+2 \sum_{n=1}^\ii \int_{0}^\ii dx e^{-|x^\al | t} \cos 2\pi n x
\label{49}
\eeq
and by comparing eqs. (\ref{46}) and (\ref{49}), we get
\beq
\Delta_\al(t)=2\sum_{n=1}^\ii \int_{0}^\ii dx e^{-|x^\al | t} \cos
2\pi n x
-\sum_{k=1}^\ii \frac{(-t)^k}{k!}\ze(-\al k).
\label{}
\eeq
The above expression can be checked immediately. In fact, for
$\al=2$, i.e. $\be=1$, one has $\ze(-2k)=0$ and
the additional term is nonvanishing, i.e.
\beq
\Delta_2(t)=2\sum_{n=1}^\ii \int_{0}^\ii dx e^{-x^2  t} \cos 2\pi
n x
= \sqrt{\frac{ \pi}{ t}}\, \sum_{n=1}^\ii  e^{-\fr{\pi^2 n^2}{t}}.
\eeq
As a consequence, eq. (\ref{46}) becomes a well known Jacobi's
theta function identity.

Furthermore, if $ \al<2$, it turns out that
$\Delta_\al(t)$ can be made very small, and we can write
\beq
2\sum_{n=1}^\ii \int_{0}^\ii dx e^{-|x^\al | t} \cos 2\pi n x
\simeq \sum_k \frac{(-t)^k}{k!}\ze(-\al k),
\label{52}
\eeq
which actually looks  quite non trivial. This sum must be cut after
its minimal term. Again, for
$\al=1$, one can perform exactly the elementary integrations
involved, and the result is
\beq
\sum_{n=1}^\ii\frac{2t}{t^2+4\pi^2n^2}=-\sum_{r=1}^\ii
\frac{t^{2r}}{(2r-1)!}\ze(1-2r).
\eeq
The sum on the left hand-side can be done with elementary
methods and we end up with
\beq
\frac{1}{2}\cot (\fr{t}{2})-\frac{1}{t}=-\sum_{r=1}^\ii
\frac{t^{2r}}{(2r-1)!}\ze(1-2r)
\eeq
which is a well known but certainly non trivial identity.

All this goes through  for $ \al = 2,4,6, \ldots $. A connection
with the theory of Brownian processes may be stablished at that
point \cite{11}. The instabilities that are known to appear outside
the range $0<\alpha <1$ and outside these
particular values of $\alpha$ can be traced back in our procedure
to the fact that both the full series in $\zeta$'s and the
additional term on the contour at infinity diverge. Then, both are
useless, in practice, and we must come back
 to our new method as stated before
(end of sect. 2) and reflected in (\ref{52}). This will
be illustrated in the following sections in great detail.

\bs
\section{Numerical estimates of the additional term}

In general, the additional term, that is, the contribution of the
semicircumference `at infinity' ---whose existence was discovered
by one of the authors a few years ago \cite{3}--- has been very
difficult to handle, even when computed at 'finite distance' Re $z=c$
or for finite radius $|z|=R$,
either analytically or numerically (see the considerations by Actor
in ref.
\cite{6}). Being such an elusive term to any kind of treatment, let us
(for the moment)
consider it in its most simple form
 \beq
 \Delta_\alpha \equiv \frac{1}{2 \pi i} \int_K dz \, \Gamma (z) \,
\zeta (\alpha z),
\eeq
where $K$ is the semicircumference of (finite) radius R, $z= R e^{i
\theta}$, with $\theta$ going from $\theta = \pi /2$ to $\theta =
3 \pi /2$. To begin with,
this expression is not defined at $\theta = \pi$ and one must start
the
calculation by using the well known reflection formulas for $\Gamma
$ and $\zeta$.
That yields
\beq
\Delta_\alpha = i \int_K \frac{dz}{ z(2\pi)^{1-\alpha z}}
\frac{\sin
(\pi  \alpha z /2)}{\sin (\pi z)} \,
 \frac{\Gamma (1-\alpha z)}{\Gamma (-z)} \, \zeta (1- \alpha z).
\label{dar}
\eeq
Such quantity is, in principle, well defined but, on the other
hand,
rather bad behaved. In modulus, the integrand clearly diverges for
$|z|
\rightarrow \infty$ but, due to the highly oscillating sinus
factors
the
final result turns out to be finite. Actually, it has been proven
rigorously in \cite{3}, that it has
the very specific value
\beq
\Delta_2= - \sqrt{\pi}  S(\pi^2), \ \ \ \ \ S(t)\equiv
\sum_{n=1}^{\infty}
e^{-\pi^2t},
\eeq
for $\alpha =2$, and that it can be made $<1$ for {\it any} value
of $\alpha >2$ and reasonably high values of $R$ (the ones actually
needed for practical applications).
A possible way to handle expression (\ref{dar}) is to employ the
integral equation
\beq
\Gamma (z) \zeta (\beta z) = \int_0^\infty S_\beta (t)  t^{z-1} dt,
\ \
\ \mbox{Re}  \ z >0, \ \ \ S_\beta (t) \equiv \sum_{n=1}^\infty
e^{-n^\beta t},
\eeq
what yields
\beq
\Delta_\alpha = -i \alpha \int_K dz \, \frac{\sin
(\pi  \alpha z /2)}{\sin (\pi z)} \,
  \frac{\Gamma (-\alpha z)}{\Gamma (-z)\Gamma (\frac{1}{\alpha}
-z)}  \int_0^\infty S_\alpha ((2\pi)^\alpha t) \
t^{\frac{1}{\alpha}
-z-1} dt.
\label{dar2}
\eeq
Using now Stirling's formula and a simple approximation for the
sinus fraction we get immediately
\beq
\Delta_\alpha = -\frac{i}{\sqrt{2\pi \alpha}} \int_K dz \,
\varphi_\alpha (z)  \int_0^\infty S_\alpha ((2\pi/ \alpha)^\alpha
t) \  t^{\frac{1}{\alpha} -z-1} dt,
\label{dar3}
\eeq
being
\bea
\varphi_\alpha (z)& \equiv & \exp \left\{ \left[ (2-\alpha) z +
\left(
\frac{1}{2} - \frac{1}{\alpha} \right)\right] \ln z + (\alpha -2)
z +  \left( \frac{\alpha}{2} - 1 \right) \pi |\mbox{Im} \ z|
\right.
\nn \\ &+& \left. i \ \mbox{sgn (Im} \, z)  \left( \frac{\alpha}{2}
- 1
\right) \pi |\mbox{Re} \ z| \right\}.
\eea
We see that, in fact, for $\alpha <1$, $\varphi_\alpha (z)
\rightarrow 0$, when $|z| \rightarrow \infty$, while for $\alpha
=2$ it is  $\varphi_2 (z) \equiv 1$ and $\Delta_2 =- \sqrt{\pi} S_2
(\pi^2)$, as anticipated before (this is nothing but the famous
Jacobi theta function identity). On the other hand, if we
substitute in (\ref{dar3}) its mean value, $\mu$, for the function
$\varphi_\alpha$, we obtain
\beq
\Delta_\alpha = -\mu \, \sqrt{ \frac{2\pi}{\alpha} } \, S_\alpha
\left( \left(\frac{2\pi}{\alpha} \right)^\alpha \right).
\label{aap}
\eeq
An analytical but approximate evaluation carried out for the
particular cases $\alpha =2N$, with $N$ a positive integer (these
values are most simple to deal with), suggests that the behavior of
the mean
value $\mu$ in terms of $\alpha$ can be bounded from above by an
expression of the kind
\beq
|\mu | \leq \frac{e-2}{2e} \sqrt{ \frac{2\pi}{\alpha} }
\eeq
for $\alpha$ large.
That gives for the additional term
\beq
|\Delta_\alpha | \leq \frac{e-2}{2e} \, S_\alpha \left(
\left(\frac{2\pi}{\alpha} \right)^\alpha \right),
\label{aap1}
\eeq
which is convergent for $\alpha \rightarrow \infty$.

We have checked this analytical bound with a numerical,
direct evaluation of the integral (\ref{dar}) written in the
following equivalent form, obtained by straightforward
manipulations,
\beq
\Delta_\alpha = - \int_{-\pi /2}^{\pi /2}
\frac{d\theta}{(2\pi)^{z+1}} \frac{\sin
(\pi z /2)}{\sin (\pi z/\alpha )} \,
 \frac{\Gamma (z+1)}{\Gamma (z/\alpha )} \, \zeta (z+1), \ \ \ z=R
e^{i \theta}, \ \ R >> 1.
\label{dar4}
\eeq
In particular, the case $\alpha \rightarrow \infty$ can be easily
handled
\beq
\Delta_\infty = -\frac{1}{\pi} \int_{-\pi /2}^{\pi /2}
\frac{d\theta}{(2\pi)^{z+1}} \, \sin
(\pi z /2)\, \Gamma (z+1) \, \zeta (z+1), \ \ \ z=R e^{i \theta},
\ \ R >> 1.
\label{dari}
\eeq
As discussed before, it is  quite  difficult to
evaluate such integrals for $R$ big. {\it Any} numerical
procedure
breaks down for $R$ large enough. This makes the method developed
in sects. 2 and 3 of this paper even more valuable. As
explained in detail there, by using this technique we {\it never}
need, in
practice, to go to very large $R$'s in the evaluation of the
additional term ---at the expense of just calculating a few first
terms of an asymptotic expansion on the zeta function of the
differential operator one is dealing with (the asymptotic series
must be optimally cut in the usual way). Put in the abstract
(simple) situation which concerns us here, we need only obtain
numerical values for the expressions (\ref{dar4}) and  (\ref{dari})
when $R$ is just of the order of 20 or 30 (what is already involved
enough!), since the usual asymptotic series that appear in practice
yield their best result for a number of terms of this (or very often
less) order.

There seems to be no definite regular behavior in the numerical
values of the expressions above, though all of them, in the
wide range $8\leq R \leq 30$, lie in the quite narrow interval
$0.02 \leq
|\Delta| \leq 0.07$. This is in very good agreement with the
results obtained from the analytical approximation (\ref{aap1}),
which are $\Delta_4 \simeq 0.04$,  $\Delta_6 \simeq 0.07$ and
$\Delta_\infty \simeq 0.13$. Only the tendency of this formula for
large $\alpha$ deviates from the numerical result by  a factor of
10 ---namely the numerical results are about ten times
lower than the upper bound values given by the formula
(\ref{aap1}). But this
is not strange, since for big $\alpha$'s this expression
(\ref{aap1}) is rather bad
($S_\alpha$ is then a very slowly convergent series), while, on the
contrary, the limiting
expression for $\Delta_\infty$ (\ref{dari}) remains in very good
shape.

Summing up, eqs. (\ref{dar4}),  (\ref{dari}) and  (\ref{aap1}) are
the best we could obtain, from an analytical point of view,
for the treatment of the
additional term which appears both in the series commutation
procedure and in the Mellin transform method.
However, for the numerical treatment, a clever use of the
methods of sect. 3 and 4
drastically reduces the problem, as we have just seen in the
present section: the calculation must be performed {\it
not} at $R=\infty$ but at reasonable values of $R$, and in this
case (\ref{dar4}) and  (\ref{dari}) can be handled through standard
numerical integration procedures.
\bs

\section{Example: exact summation of the string partition
function for different ranges of temperature}

The problem of deciding if QCD is actually (some limit of) a
theory of extended objects (strings, membranes or $p$-branes in
general) is almost twenty years old and goes back to 't Hooft. In
a
very recent contribution, Polchinski has tried to answer this
question both for the Nambu-Goto and for the rigid strings by
calculating their partition functions and seeing if they actually
match with the one corresponding to QCD for different ranges of
temperature \cite{9}. This is no place to go into the details of the
procedure \cite{9}, which will be given in a forthcoming paper and we
shall here concentrate in the mathematical aspects of the problem
only.

The first two terms in the loop expansion
\beq
S_{eff} = S_0+S_1+ \cdots
\eeq
 of the effective action corresponding to the rigid string
\beq
S= \frac{1}{2\alpha_0} \int d^2\sigma \, \left[ \rho^{-1}
\partial^2 X^\mu \partial^2 X_\mu + \lambda^{ab} \left( \partial_a
X^\mu \partial_b X_\mu - \rho \delta_{ab} \right) \right] + \mu_0
 \int d^2\sigma \, \rho,
\eeq
being $\alpha_0$ the dimensionless, asymptotically free coupling,
$\rho$ the intrinsic metric, $\mu_0$ the explicit string tension
(important at low energy) and $\lambda^{ab}$ the Lagrange
multipliers, are given, in the world sheet $0\leq \sigma^1 \leq L$
and $0\leq \sigma^2 \leq \beta t$ (an annulus of modulus $t$),  by
\beq
S_0= \frac{L\beta t}{2\alpha_0} \left[ \lambda^{11} +  \lambda^{22}
t^{-2} + \rho \left( 2\alpha_0 \mu_0 -  \lambda^{aa} \right)
\right]
\eeq
(tree level) and
\bea
S_1 &=& \frac{d-2}{2} \, \ln \det \left( \partial^4- \rho
\lambda^{ab} \partial_a  \partial_b \right) \nn \\
&=&  \frac{d-2}{2} \, L \sum_{n=-\infty}^{\infty} \int_{-
\infty}^{+\infty} \frac{dk}{2\pi} \, \ln \left[ \left( k^2 +
\frac{4\pi^2n^2}{\beta^2t^2} \right)^2 + \rho \left(  \lambda^{11}
k^2 + \frac{4\pi^2n^2}{\beta^2t^2 \lambda^{22}} \right) \right]
\eea
(one loop). This is a highly non-trivial calculation, which has
been performed in ref. \cite{9} only in the very strict limits
$T\rightarrow 0$ and $T\rightarrow \infty$ around some extremizing
configuration, the parameters being $\rho$, $\lambda^{11}$,
$\lambda^{22}$ and $t$.
As a first step of the zeta function method we write
\beq
S_1= -(d-2) \left. \frac{d}{ds} \zeta_A (s/2)\right|_{s=0}, \
\ \ \ \zeta_A (s/2) = L  \sum_{n=-\infty}^{\infty} \int_{-
\infty}^{+\infty} \frac{dk}{2\pi} \, (k^2+y_+^2)^{-s/2}  (k^2+
y_-^2)^{-s/2},
\eeq
where
\beq
y_{\pm} = \frac{a}{t} \left[ n^2 + \frac{\rho
t^2\lambda^{11}}{2a^2} - \sqrt{\rho} \, \frac{t}{a} \left(
(\lambda^{11}-\lambda^{22}) n^2 \pm \frac{\rho
t^2\lambda^{11^2}}{4a^2} \right)^{1/2} \right]^{1/2}, \ \ \ a\equiv
\frac{2\pi}{\beta}.
\eeq
One may consider two different approximations of overlapping
validity \cite{9}, one for low temperature, $\beta^{-2} << \mu_0$, and
the other for high temperature,  $\beta^{-2} >> \alpha_0\mu_0$.
Both these approximations (overlapping included) can be obtained
from the exact expression above. It can be written in the form
\beq
 \zeta_A (s/2) = \frac{L}{2\sqrt{\pi}} \, \frac{\Gamma (s-
1/2)}{\Gamma (s)}\,  \sum_{n=-\infty}^{\infty} \frac{y_-
}{(y_+y_-)^s} \, F(s/2,1/2;s; 1-\eta), \ \ \ \ \eta\equiv \frac{y_-
^2}{y_+^2}.
\label{zz1}
\eeq

This is an exact formula.
For high temperature, the ordinary expansion of the confluent
hypergeometric function $F$ is in order
\beq
F(s/2,1/2;s; 1-\eta) = \sum_{k=0}^{\infty}
\frac{(s/2)_k(1/2)_k}{k!(s)_k} \, (1-\eta )^k \, \longrightarrow
1 + \eta^{-1/2}, \ s \rightarrow 0,
\eeq
since $y_\pm$ can be written as
\beq
y_\pm \simeq  \frac{a}{t} \,   \left[ (n
\pm b)^2+c^2 \right]^{1/2}, \ \ \ \   b=\frac{\beta t}{4\pi}
\sqrt{\rho
(\lambda^{11}- \lambda^{22} )}, \ \ c\simeq \frac{\alpha_0}{4\pi}.
\label{yes}
\eeq
After the appropriate analytic continuation, the derivative of the
zeta function yields
\beq
 \left. \frac{d}{ds} \zeta_A (s/2)\right|_{s=0}
= -\frac{L}{4} \sqrt{b^2+c^2} +\frac{2\pi L}{\beta t} \left[ b^2+
\frac{1}{6} - \left( \frac{1}{2} \ln (-b^2) + \psi (-1/2) + \gamma
\right) c^2 \right].
\eeq
In order to obtain this result, which comes from elementary Hurwitz
zeta functions, $\zeta (\pm 1,b)$, we have used the binomial
expansion in (\ref{yes}) what is
completely consistent with the approximation (notice the extra
terms coming from the contribution of the pole of (\ref{yes}) to
the derivative of $\zeta_A (s/2)$).

The low temperature case is more involved and now the
methods of sect. 3 prove to be very useful. The term $n=0$
 must be treated separately. It gives
\beq
\zeta_A^{(n=0)} (s/2) = \frac{L}{2\pi} \, \frac{\Gamma ((1-s)/2)
\Gamma(s-1/2)}{\Gamma (s/2)} \, (\lambda^{11} \rho )^{1/2-s}, \ \
\ \ \frac{1}{2} < \Re (s) < 1.
\eeq
This is again an exact result, which yields
\beq
 \left. \frac{d}{ds} \zeta_A^{n=0} (s/2)\right|_{s=0} =-
\frac{L}{2} \, \sqrt{\lambda \rho}
\eeq
and
\beq
S_1^{n=0} = (d-2) \frac{L}{2} \, \sqrt{\lambda \rho}.
\label{s10}
\eeq
Such contribution must be added to the one coming from the
remaining
terms  (eq. (\ref{zz1}) above)
\beq
 \zeta_A' (s/2) = \frac{L}{\sqrt{\pi}} \, \frac{\Gamma (s-
1/2)}{\Gamma (s)}\,  \sum_{n=1}^{\infty} \frac{y_-
}{(y_+y_-)^s} \, F(s/2,1/2;s; 1-\eta), \ \ \ \
\Re (s) >1.
\label{zp1}
\eeq
(The prime is no derivative, it just means that the term $n=0$ is
absent here.) \, Within this approximation, and working around the
classical, $T=0$ solution: $\lambda^{11}$ $ =\lambda^{22} $ $
=\alpha_0 \mu_0$, $\rho = t^{-2} =1$, we obtain
\beq
 \zeta_A' (s/2) = \frac{L}{\sqrt{\pi}} \, \frac{\Gamma (s-
1/2)}{\Gamma (s)}\, \left(\frac{a}{t} \right)^{1-2s} \left[ F_0
(s)+F_1 (s)+F_2 (s)  \right],
\label{zp2}
\eeq
where
\bea
&& F_0 (s) \equiv  \sum_{n=1}^{\infty} \frac{n^{1-
s}}{(n^2+\sigma^2)^{s/2}} \,
\left[F(s/2,1/2;s;\sigma^2/(n^2+\sigma^2)) -1 -
\frac{\sigma^2}{4n^2} \right], \ \ \ \sigma^2\equiv \frac{\lambda
\rho t^2\beta^2}{4\pi^2}, \nn \\
&& F_1 (s) \equiv  \sum_{n=1}^{\infty} \frac{n^{1-
s}}{(n^2+\sigma^2)^{s/2}}, \ \ \ \ F_2 (s) \equiv
\frac{\sigma^2}{4} \, \sum_{n=1}^{\infty} \frac{n^{-
1-s}}{(n^2+\sigma^2)^{s/2}}.
\label{efes}
\eea

The study of these functions  is done in appendix A. It involves
the
procedures of sect. 3 and is very related with example 2 there.
Substituting the results of the appendix back into  (\ref{zp2}), we
obtain
\bea
 \zeta_A' (s/2) & = & \frac{L}{\sqrt{\pi}} \, \frac{\Gamma (s-
1/2)}{\Gamma (s)}\, \left(\frac{a}{t} \right)^{1-2s} \left[
\frac{\sigma^2}{8s} -\frac{\sigma^2}{8} - \frac{\sigma^2}{4} \ln
\left(  \frac{\sigma}{2} \right) -\frac{\sigma}{4} \right. \nn \\
&-& \left. \frac{1}{24} +  \frac{1}{2\pi}  \int_\sigma^{\infty} dr
\, \ln \left( 1- e^{2\pi r} \right) \frac{r}{\sqrt{r^2-\sigma^2}}
+ {\cal O} (s)\right],
\eea
therefore
\beq
 \zeta_A' (0) = -\frac{L a\sigma^2}{4t}
\eeq
and
\bea
 \left. \frac{d}{ds} \zeta_A' (s/2)\right|_{s=0} &=& - \frac{La}{t}
\left[ \psi (-1/2) \frac{\sigma^2}{4}-  \ln (a/t)
\frac{\sigma^2}{2} - \frac{\sigma^2}{4} - \frac{\sigma^2}{2} \ln
(\sigma /2) -\frac{\sigma}{2} \right. \nn \\ &+& \left.  \gamma \,
\frac{\sigma^2}{4}  -
\frac{1}{12} +  \frac{1}{2\pi}  \int_\sigma^{\infty} dr \, \ln
\left( 1- e^{2\pi r} \right) \frac{r}{\sqrt{r^2-\sigma^2}} \right].
\eea
Finally
\bea
S_1^{(n\neq 0)}  &=& (d-2) \frac{La}{t} \left[ \psi (-1/2)
\frac{\sigma^2}{4}-  \ln (a/t) \frac{\sigma^2}{2} -
\frac{\sigma^2}{4} - \frac{\sigma^2}{2} \ln (\sigma /2) -
\frac{\sigma}{2} \right. \nn \\ &+& \left.  \gamma - \frac{1}{4} +
\frac{1}{2\pi}  \int_\sigma^{\infty} dr \, \ln \left( 1- e^{2\pi r}
\right) \frac{r}{\sqrt{r^2-\sigma^2}} \right].
\label{s1n0}
\eea
Adding (\ref{s10}) and (\ref{s1n0}) we obtain the desired
expansion  of the one loop effective action, $S_1$, near $T=0$.
The physical implications of these
results will be discussed elsewhere.
\bs

\section{Discussion and conclusions}

The series commutation techniques that are essential in the proof
of the zeta function regularization theorem ---which, on its turn,
is the basic tool in the general procedure of zeta function
regularization when the spectrum of the operator is explicitly
known--- have been promoted in this paper to an extremely elegant
mathematical method, by making use of the Mellin transforms of
convenient heat kernel operators, in
combination with a rigorous treatment of the asymptotic series
involved. The laborious analysis of the series to be commuted,
 the artificial picking up of a convenient function
in order to mimick such series through pole residues on the complex
plane, and the process of
commutation itself, with the appearance of additional terms `at
infinity', have
now (almost completly) disappeared, in favor of a quite natural
Mellin transform
(better, inverse Mellin transform) analysis of the heat kernels
(see also \cite{8}). Moreover,
the identification of
the three  different contributions to the final result \cite{6},
namely, the naive commuted series (which results in a sum of
Riemann or Hurwitz zeta functions), the ordinary commutation
remnants (a polynomial function) and the very elusive,
additional term of negative power-like behavior, appears now in a
clear, natural, almost misterious way. The last contribution \cite{3},
which was originally very difficult to handle from a numerical
point of
view, has been given here a completely new treatment, which allows
the calculation of explicit numbers with reasonable ease.

Our new method has the additional advantage that it is equally well
fitted for the treatment of general elliptic differential operators
whose
spectrum is {\it not} known (notice that the procedures in \cite{2,3,6}
depend heavily on the explicit knowledge of the full spectrum).
 The sum over eigenvalues is then
naturally
replaced ---within the same procedure--- by a sum over heat-kernel
or Seeley-De Witt coefficients. A huge
mathematical industry has been generated for the calculation of
these coefficients, and one can now get full profit from these
result
in the new context of the zeta function regularization theorem.

 The physical applications one can envisage for the techniques here
developed keep
growing every day. Aside from the many calculations, carried out in
different contexts, of the
vacuum energy and the Casimir effect in quantum field
theory, condensed matter and solid state physics \cite{10}, in the last
section we have already hinted at the straightforward use one
can make of our expressions in the calculation of the partition
functions of strings and membranes. As pointed out  by
Polchinski et al. \cite{9}, these results are essential in order to
decide, once and forever, if QCD can possibly be a certain limit of some
string
(or membrane, or p-brane) theory, through the analysis of the high
temperature behavior of the corresponding partition functions.
These kind of calculations are considered to be very difficult
even by the mathematical-physics community (see the specific
description in \cite{9}). As we have proven in the last section of this
paper,
they can be handled in a relatively simple way through our methods.
We hope to be able to  elaborate further on these and other
applications in the near future.

\vspace{5mm}

\ni{\large \bf Acknowledgments}

This work has been carried out under the auspices of a CICYT-INFN
spanish-italian collaboration. Partial finantial support from
Generalitat de Catalunya is also acknowledged. We are very grateful
to Klaus Kirsten for a critical reading of the manuscript and
for pointing us ref. \cite{8}.
\bs

\appendix
\section{Appendix: Auxiliar expressions for the string partition
function at low temperature}

We make here a carefull study of the functions $F_i$ defined in
(\ref{efes}).
$F_1$ and $F_2$ can be analytically continued without problems:
\beq
 F_1 (s) =  \zeta (2s-1) - \frac{s\sigma^2}{2} \zeta (2+1) +
\sum_{n=1}^{\infty} n^{1-
s} \left[ (n^2+\sigma^2)^{-s/2} -n^{-s} + \frac{s\sigma^2}{2} \,
n^{-s-2} \right],
\eeq
so that, in particular,
\beq
F_1 (0) = -\frac{1}{12} - \frac{\sigma^2}{4},
\eeq
and
\beq
F_2(s)=  \frac{\sigma^2}{4} \,  \zeta (2+1) +  \frac{\sigma^2}{4}
\sum_{n=1}^{\infty} n^{-1-s} \left[ (n^2+\sigma^2)^{-s/2} -n^{-s}
\right],
\eeq
which has a pole at $s=0$,
\beq
F_2(s)=  \frac{\sigma^2}{8s}+  \frac{\gamma \sigma^2}{4}+ {\cal O}
(s).
\eeq
Concerning $F_0$, by using
\beq
\lim_{s\rightarrow 0} F(s/2,1/2;s;\sigma^2/(n^2+\sigma^2)) =
\frac{1}{2} \left[ 1+ \sqrt{ 1 + \left( \frac{\sigma}{n} \right)^2}
\right],
\eeq
we can write
\beq
F_0 (0) =  \frac{1}{2} \sum_{n=1}^{\infty} \left[
(n^2+\sigma^2)^{1/2} -n - \frac{\sigma^2}{2n^2} \right] =
\frac{1}{2}\, \lim_{\tau \rightarrow 0} \sum_{n=1}^{\infty} \left[
\left( (n^2+\sigma^2)^{1/2} -n - \frac{\sigma^2}{2n^2} \right) e^{-
\tau n} \right].
\eeq
This yields
\beq
 \sum_{n=1}^{\infty}  n \,  e^{-\tau n} = \frac{1}{\tau^2} -
\frac{1}{12} + {\cal O} (1),
\eeq
for the middle term. For the other two we shall make explicit use
of the results obtained in sect. 3 of this paper. For the last
term, we get
\beq
 \sum_{n=1}^{\infty}  \frac{1}{n} \,  e^{-\tau n} =
\mbox{Res}_{z=0} \, \left[ \tau^{-z} \Gamma (z) \zeta (z+1) \right]
+ {\cal O} (1) = - \ln \tau + {\cal O} (1).
\eeq
In order to apply this procedure to the first term we need some
more specific knowledge of the function
\beq
\zeta_{-1/2} (z) \equiv  \sum_{n=1}^{\infty} n^{-z} \,
(n^2+\sigma^2)^{1/2},
\eeq
which appears when considering (see sect. 3)
\bea
 \sum_{n=1}^{\infty}   (n^2+\sigma^2)^{1/2} \, f(\tau n) &=&
 \sum_{n=1}^{\infty}   (n^2+\sigma^2)^{1/2} \frac{1}{2\pi i} \int
dz \, (n \tau )^{-z} M[f] (z) \nn \\ &=&   \frac{1}{2\pi i} \int dz
\,
\tau^{-z} M[f] (z) \, \zeta_{-1/2} (z).
\label{mte1}
\eea
The study of this function can be reduced to example 2 of sect. 3.
For Re $z > -2p$ we get an analytical continuation which is a
meromorphic function with poles  at $z=2(1-k)$ of residues $\left(
{1/2 \atop k} \right) \sigma^{2k}$, $k=0,1,2,3, \ldots$
\bea
\zeta_{-1/2}^\epsilon (z) &=& \int_{\epsilon}^{\infty} dr \, r^{1-
z} \left[ \sqrt{1+ \frac{\sigma^2}{r^2} } - \theta (p-1)
\sum_{k=0}^p \left( {1/2 \atop k} \right) \left( \frac{\sigma}{r}
\right)^{2k}  \right] \nn \\
&+&  \sum_{k=0}^p \left( {1/2 \atop k} \right)
\frac{\sigma^{2k}}{z+ 2(k-1)} +  \epsilon^{1-z}
(\epsilon^2+\sigma^2)^{1/2}  - \frac{1}{2}  \epsilon^{-z}
(\epsilon^2+\sigma^2)^{1/2} \nn \\
&-&  \frac{i}{2\pi} \sum_{\pm} \int_{C^{(\pm)}} dt \, \ln \left( 1-
e^{\pm 2\pi i t} \right) \frac{d}{dt} \left[ t^{-z}
(t^2+\sigma^2)^{1/2} \right],
\eea
where the three last terms are integer functions and the contours
$C^{(\pm)}$ are chosen avoiding the points $\pm i \sigma$.

In our case it is sufficient to take $p=1$. Considering now $-2 <$
Re $z <0$, the limit $\epsilon \rightarrow 0$ can be taken
naively, with the result
\bea
\zeta_{-1/2} (z) &=& \int_0^{\infty} dr \, r^{1-z} \left[ \sqrt{1+
\frac{\sigma^2}{r^2} } -\theta (p-1) \sum_{k=0}^p \left( {1/2 \atop
k} \right) \left( \frac{\sigma}{r} \right)^{2k} \right] \\
&+&  \sum_{k=0}^1 \left( {1/2 \atop k} \right)
\frac{\sigma^{2k}}{z+ 2(k-1)} - \frac{i}{2\pi} \sum_{\pm}
\int_{C^{(\pm)}} dt \, \ln \left( 1- e^{\pm 2\pi i t} \right)
\frac{d}{dt} \left[ t^{-z} (t^2+\sigma^2)^{1/2} \right]. \nn
\eea
 It is the last term the one which prevents the continuation to the
right of $z=0$ due to its divergence at $t=0$. This needs a special
treatment (see again sect. 3), and the final result is
\beq
\zeta_{-1/2} (z) = \frac{\sigma^2}{4} - \frac{\sigma^2}{2} \ln
\left(  \frac{\sigma}{2} \right) -  \frac{\sigma}{2}
+\frac{\sigma^2}{2z} +  \frac{1}{\pi}  \int_\sigma^{\infty} dr \,
\ln \left( 1- e^{2\pi r} \right) \frac{r}{\sqrt{r^2-\sigma^2}} +
{\cal O} (z).
\eeq
The integral term is awkward but harmless: it is exponentially
suppressed as compared to the rest. Now we can go back to
(\ref{mte1})
\bea
 && \sum_{n=1}^{\infty}   (n^2+\sigma^2)^{1/2} \, e^{-\tau n} =
  \frac{1}{2\pi i} \int dz \,  \tau^{-z} \Gamma (z) \, \zeta_{-1/2}
(z) = \frac{1}{\tau^2} -  \frac{\sigma^2}{2} \ln \tau  \\
& & - \frac{\gamma\sigma^2}{2} + \frac{\sigma^2}{4} -
\frac{\sigma^2}{2} \ln \left(  \frac{\sigma}{2} \right) -
\frac{\sigma}{2} +  \frac{1}{\pi}  \int_\sigma^{\infty} dr \, \ln
\left( 1- e^{2\pi r} \right) \frac{r}{\sqrt{r^2-\sigma^2}} + {\cal
O}_\tau (1). \nn
\eea
And this yields for $F_0(0)$:
\beq
F_0(0) =  - \frac{\gamma\sigma^2}{4} + \frac{\sigma^2}{8} -
\frac{\sigma^2}{4} \ln \left(  \frac{\sigma}{2} \right) -
\frac{\sigma}{4} +\frac{1}{24} +  \frac{1}{2\pi}
\int_\sigma^{\infty} dr \, \ln \left( 1- e^{2\pi r} \right)
\frac{r}{\sqrt{r^2-\sigma^2}}.
\eeq

\newpage



\end{document}